\documentclass[a4paper,12pt]{article}
\usepackage{amsmath}
\usepackage{latexsym}
\def\drm{{\rm d}}
\def\pd#1#2{\frac{\mbox{$\partial #1$}}{\mbox{$\partial #2$}}}
\def\pdel#1#2{\frac{\mbox{$\delta #1$}}{\mbox{$\delta #2$}}}
\def\pdelu#1#2{\frac{\mbox{$\delta #1$}}{\mbox{$\delta u(x#2)$}}}
\def\pcD#1#2{\frac{\tsl \cD#1}{\tsl \cD#2}}
\def\dd#1#2{\frac{\tsl \drm#1}{\tsl \drm#2}}
\def\pcDb#1#2{\frac{\tsl \cDb#1}{\tsl \cDb#2}}
\def\vx{{\bf{x}}}
\def\vu{{\bf{u}}}
\def\dx{{\rm d \vx}}
\def\tsl{\textstyle}
\def\vPhi{\bar{\Phi}}  
\def\tb{{\bar t}}
\def\ub{{\bar u}}
\def\vub{{\bar{\vu}}}
\def\vphi{{\boldsymbol \phi}}
\def\cD{{\cal D}}
\def\vxi{{\boldsymbol \xi}}
\def\cDb{{\bar\cD}}
\def\Xrm{{\rm X}}
\def\e{\varepsilon}

\begin{document}

\title{
Application of Lie group analysis to functional
differential equations}
\date{}
\maketitle
\noindent
\begin{center}
Martin Oberlack$^1$ and Marta Wac\l{}awczyk$^2$ \\
1.  Depertment of Mechanical Engineering, Faculty of Fluid Dynamics  \\
    Darmstadt University of Technology, \\
    Petersenstra{\ss}e 13, 64287, Darmstadt, Germany \\
2.   Institute of Fluid Flow Machinery, \\  
    Polish Academy of Sciences, 
    ul.\ Fiszera 14, 80952 Gda\'nsk, Poland. 
\end{center}

\begin{abstract}
In the present paper the classical point symmetry
analysis is extended from partial differential
to functional differential equations with functional
derivatives. In order to perform the group analysis
and deal with the functional derivatives 
we extend
the quantities such as infinitesimal transformations,
prolongations and invariant solutions.
For the sake of example
the procedure is applied to the continuum limit of the
heat equation. 
The method  can further lead to significant applications
in statistical physics and fluid dynamics.


\end{abstract}


\section{Introduction}
In the paper we consider such functional differential equations
which can be regarded as the extensions of partial differential
equations.
The key idea is that the  discrete set of 
independent variables
$(y_1, y_2, \dots, y_n)$ in a partial equation
is replaced by a continuous set of infinitely many
variables denoted by $[y(x)]$. 
In order to illustrate the above extension for partial differential 
equations to
functional differential equations we introduce the example
\begin{equation}\label{ExamplePDE}
\pd{f}{t} = \sum_{i=1}^n y_i \pd{f}{y_i}
\end{equation}
where $f = f(t;y_1, y_2, \dots, y_n)$.
Taking the continuum limit we obtain
\begin{equation}\label{ExampleFDE}
\pd{f}{t} = \int_l y(x) \frac{\delta f}{\delta y(x)} \, \drm x
\end{equation}
where $f = f(t;[y(x)])$ and the partial derivatives in (\ref{ExamplePDE})
have been replaced by a functional derivative $\delta / \delta y(x)$
which can also be denoted by $\partial/\partial y(x)\drm x$,
(cf.\ Gelfand \& Fomin (1963) for the definition of the 
functional derivative).
The latter notation is more convenient in some situations (Hopf, 1952).
The similar procedure of deriving the functional differential equations
as the limit case of partial differential equations 
is used also in the work of Hopf (1952) and Breuer \& Petruccione (1994).

Functional equations are used to describe problems of 
statistical mechanics where the probability distributions of phases and their
time evolution are studied. In a discrete medium 
the phase space at a given 
time instant $t$ is defined by
velocities and positions of all particles
$(\vx_1,\vx_2,\dots,\vx_n,\vu_1,\vu_2,\dots,\vu_n)$
 contained in a considered domain.
 In a continuum limit (eg.\ in hydromechanics)
the phase becomes a continuous function of
spatial variables $\vu=\vu(\vx)$. The time evolution
of its probability distribution is  determined by a
functional equation. 

Many examples of the functional equations with
functional derivatives can be found
in physics. The areas of interests of the authors are the 
turbulent flows and turbulent reacting flows. The 
idea of description of turbulence in terms of the 
characterisctic functional has been introduced in 
a seminal work by E.\ Hopf (1952).  The approach is presented 
in the book of Monin \& Yaglom (1971), 
we mention also the work of Lewis \& Kraichnan (1962).
The similar functional formulation can also be derived 
for the Burgers equation. Functional formulation of 
Burgers equation is studied in the works of 
Breuer \& Petruccione (1992) and Breuer et al (1996). 
The authors perform stochastic simulation of the Burgers 
model. The method is based on a discrete master 
equation which is equivalent to the Hopf functional 
equation in the limit of continuous space.
The approach has been continued in works of 
Breuer \& Petruccione (1994),  Biechele et al. (1999),
Friedrich (2002). 
Another functional differential equation is used for 
the statistical description of a turbulent premixed flame 
(see Oberlack etal. 2001). 

The approach of Hopf originates from the field theory 
where the functional equations and functional derivatives 
are also used (cf. Itzykson \& Drouffe 1989, 
 Schweber 1962).
The examples of functional 
equations in field theory are the Schwinger equations 
(cf. Pester et al. 2002) and Wheeler de Witt 
equations (cf. DeWitt 1967, Barvinsky \& Kiefer 1998). 


The symmetry analysis based on the Lie group
theory has become a powerful tool of analysing,
simplifying and finding solutions of partial differential
equations (cf.\ Ibragimov 1994, 1995, 1996, Cantwell, 2002).  
The method also gives a deep insight into the 
underlying physical problems described by the
differential equation.  
Examples of its applications include
problems of fluid dynamics, where 
 a broad range of invariant solutions for turbulence
statistics were found (Oberlack, 1999, 2001).
However, much less attention have been given so far to
the symmetry analysis of functional equations. 
Some of the previous works concern the integro-differential 
(Zawistowski 2001, Chetverikov \& Kudryavtsev 1995, Roberts 1985) 
and  delay differential equations 
(Zawistowski 2002, Tanthanuch \& Meleshko, 2004).
However, to the best of the authors knowledge,
the type of functional equations considered in the present
 paper, has not been studied so far in terms of
the symmetry analysis.
In the present work the classical symmetry analysis is extended 
to functional differential equations which contain functional derivatives. 
The structure of the paper is the following:
first, we introduce the necessary notation to study functional
differential equations. In the main section we extend classical
Lie group methods to functional differential equations by
extending quantities such as infinitesimal transformations, prolongations or
invariant solutions. Finally, as an example, the new method is
 applied to
a continuum limit of the heat equation. This leads to the 
 transformation groups  as well as
to invariant solutions of the considered 
functional equation.

\section{Notations} \label{SecNotation}

In the present work we study the extension of the  partial differential
equation for a scalar function $\Phi(u_1,\dots, u_n,t_1,\dots, t_m)$
of $m+n$ independent variables. The general form of the
differential equation describing $\Phi$ writes:
\begin{equation}\label{pde}
  F(u_1,\dots, u_n,t_1,\dots, t_m,\Phi,\underset{1}{\Phi},
\underset{2}{\Phi},\dots,\underset{q}{\Phi})=0
\end{equation}
where $\underset{k}{\Phi}$ denote the $k$-th derivatives of the function 
$\Phi$ with respect to any possible combination of independent variables
and $q$ is the highest order of derivative present in Eq.~\ref{pde}.
In the considered continuum limit $u$ becomes a 
function of the continuous 
variable $\vx$. For the sake of generality we
consider that $\vu=\vu(\vx)$ where $\vu=(u_1,u_2,\dots,u_p)$
and $\vx=(x_1,x_2,\dots,x_r)$ are vectors in a $p$- and 
$r$-dimensional space.
In this case $\Phi$ becomes a functional 
\begin{equation}
\Phi=
\Phi([u_1(\vx)],\dots,[u_p(\vx)],t_1,\dots,t_m)
=\Phi([\vu(\vx)],t_1,\dots, t_m)
\end{equation}
and the partial
differential equation (\ref{pde}) becomes a functional differential
equation:
\begin{equation}\label{fde}
 F([\vu(\vx)],t_1,\dots, t_m,\Phi,
\underset{1}{\Phi},
\underset{2}{\Phi},\dots,\underset{q}{\Phi})=0;
\end{equation}
here again, $\underset{k}{\Phi}$ denotes all possible derivatives
of order $k$, which can include partial derivatives with respect
to $t_i$ and functional derivatives with respect to $u_\alpha(\vx)$. 
The following, equivalent notation will
be used for the first functional derivatives:
\begin{equation}\label{pd}
\Phi_{,u_\alpha(\vx)}=\pdel{\Phi}{u_\alpha(\vx)}=\pd{\Phi}{u_\alpha(\vx)\dx}.
\end{equation}
The different types of second order derivatives in Eq.~(\ref{fde})
will be  denoted by
\begin{eqnarray}
\Phi_{,t_i t_j}=\pd{^2\Phi}{t_i \, \partial t_j}, \quad
\Phi_{,u_\alpha(\vx) t_j}=\pd{^2\Phi}{u_\alpha(\vx)\dx \, \partial t_j}, 
\quad \Phi_{,u_\alpha(\vx) u_\beta(\vx')}
=\pdel{^2\Phi}{u_\alpha(\vx) \delta u_\beta(\vx') },
\nonumber
\end{eqnarray}
while the order of any derivative is commutative.
Higher order derivatives can be expressed in an analogous way.

As $\vx$ is a vector in a $r$-dimensional space,
we will use the following, convenient notation for
 the integrals with respect to $\vx$: 
\begin{eqnarray}
\int_{\bf G} \dx = 
\underbrace{\int\int\dots\int}_{\bf G}\drm x_1 \drm x_2 \dots \drm x_r 
\end{eqnarray}
where ${\bf G}$ is a domain in $r$-space to be specified
for each particular problem.

For further purposes we recall here some differentiation and
integration rules for functionals. We also solve a simple 
functional equation by the method of characteristics.
For the sake of clarity we will first present 
necessary formulae for the partial differential
equation (\ref{pde}) and introduce their counterparts
in the continuum limit (\ref{fde}).
The two approaches will also be called ``classical''
and ``continuum formulation'', respectively.

A function  $\Phi(u_1,\dots, u_n,t_1,\dots, t_m)$ of
a finite set of variables is differentiable for a particular
value of its arguments if its variational (or functional) form
$\delta \Phi$ is linear in $\delta u_i$, i.e.\ the
following relation holds (cf.\ Hopf, 1952)
\begin{equation}
\delta \Phi = \sum_{i=1}^{n}\pd{\Phi}{ u_i} \delta u_i
+ \sum_{i=1}^{m}\pd{\Phi}{t_i}\delta t_i.
\end{equation}
The analogous formula in the considered continuum limit writes
\begin{equation}\label{dif_int}
\delta \Phi= \sum_{\alpha=1}^{p} \int_{\bf G}\pdel{\Phi}{ u_\alpha(\vx)} 
\delta u_\alpha(\vx) \dx
+ \sum_{i=1}^{m}\pd{\Phi}{t_i}\delta t_i.
\end{equation}

A functional derivative of $\Phi$ exists if its differential
form $\delta \Phi$ can be written as the integral (\ref{dif_int}).
Let us consider the following function of a finite number of variables
$\Phi = \sum_{i=1}^{n} a_i u_i.$
Its derivative with respect to $u_k$ where $1 \le k \le n$ is
$\partial \Phi/\partial u_k= 
 \sum_{i=1}^{n} a_i \delta_{ik} = a_k$
where $\delta_{ik}$ is a Kronecker delta. 
We consider the following continuum limit
of the function $\Phi$
\begin{equation}
\Phi = \int_{G} a(x') u(x') \drm x'.
\end{equation}
The differential form (\ref{dif_int}) of this functional writes:
\begin{equation}
\delta \Phi = \int_{G} a(x') \delta u(x') \drm x'.
\end{equation}
Hence, by comparison to  (\ref{dif_int}) the functional derivative
$\delta \Phi / \delta u(x) = a(x)$ where $x\,\epsilon\,G $. 
Another functional
$\Phi=u(x_1)$ where $x_1$ is a given point
in the domain $G$, can also be written in 
the integral form
\begin{equation}
\Phi = \int_{G} \delta(x-x_1) u(x) \drm x \quad \textrm{and} \quad
\delta \Phi = \int_{G} \delta(x-x_1) \delta u(x) \drm x,
\end{equation}
also in this case the functional derivative can be found 
 by comparison with (\ref{dif_int}), 
 $\delta \Phi/\delta u(x)=\delta(x-x_1)$.
Calculation of second order derivatives is presented on another
example below. We consider the following
function, together with its continuum counterpart
\begin{equation}\label{}
\Phi = \sum_{i=1}^{n}\sum_{j=1}^n c_{ij} u_i u_j \quad \to \quad
\Phi = \int_{G} \int_{G} c(x,x') u(x') u(x) \,  \drm x\, \drm x'.
\end{equation} 
Corresponding derivatives of the first and second order  write
\begin{eqnarray}\label{sec_der}
\pd{\Phi}{u_k} =  \sum_{j=1}^n c_{kj} u_j  +  \sum_{i=1}^n c_{ik} u_i  
\quad \to \quad
\pdel{\Phi}{u(x'')} &=& 
\int_{G} c(x'',x') u(x') \, \drm x' \\
&+& 
\int_{G} c(x,x'')  u(x) \, \drm x.  \nonumber \\
 \pd{^2\Phi}{u_k \partial u_m} = c_{km} + c_{mk} \quad \to \quad 
\pdel{^2\Phi}{u(x'') \delta u(x)} &=&  c(x'',x) +  c(x,x''). 
\end{eqnarray}
Derivatives of higher orders of the considered function and the
corresponding functional are zero.

We now solve the following hyperbolic
equation in a classical and continuum
formulation by the method of characteristics
\begin{equation}\label{eq_char}
\Phi\pd{F}{\Phi}+\sum_{i=1}^n\pd{F}{u_i}=0  \quad \to \quad
\Phi\pd{F}{\Phi} + \int_{a}^b \pdel{F}{u(x')} \, \drm x' = 0
\end{equation}
where $F=F(\Phi,u_1,\dots,u_n)$ in the classical formulation and
$F=F(\Phi,[u(x)])$ in the continuum limit; $u_i$ and $u(x)$
constitute sets of independent variables and $\Phi$ is a
dependent variable $\Phi=\Phi(u_1,\dots,u_n)$ or
$\Phi=\Phi([u(x)])$.
The characteristic equations of (\ref{eq_char}) 
\begin{equation}\label{sys_char}
\frac{\drm \Phi}{\Phi}=\drm u_1=\dots=\drm u_n
 \quad \to \quad
\frac{\delta \Phi}{\Phi}= \delta u(x) 
\quad \textrm{for each} \quad x\,\epsilon\,(a,b)
\end{equation}
determine $n$ integration constants $C_i$ in the classical formulation
and an infinite set of integration constants $C(x)$ in the continuum 
formulation. The constants can be employed as new (dependent and independent)
variables of $F$. Corresponding solutions of Eqs (\ref{eq_char})
have the forms $F=F(C_1,\dots,C_n)$ and $F=F(C_1,[C(x)])$.
A few examples of possible solutions of the characteristic system
(\ref{sys_char}) are presented below. We can e.g.\ consider
equations
\begin{eqnarray}
 \frac{\drm \Phi}{\Phi}= \drm u_1,\qquad
\drm u_1=\drm u_2,\qquad \drm u_2=\drm u_3, \quad \dots, \quad
\drm u_{n-1}=\drm u_n \nonumber \\
 \textrm{or} \qquad
\frac{\drm \Phi}{\Phi}= \frac{1}{n} \sum_{i=1}^n \drm u_i,\qquad
\drm u_1=\drm u_2,\qquad \drm u_1=\drm u_3,\quad \dots, \quad
\drm u_{1}=\drm u_n \nonumber
\end{eqnarray}
to obtain the following integration constants
\begin{eqnarray}\label{sol_char}
C_1=\frac{\Phi}{\exp{u_1}}
\qquad \textrm{or} \qquad 
C_1= \Phi \exp{\left(-\frac{1}{n}\sum_{i=1}^n u_i\right)},
 \nonumber \\
\;\; C_2=u_1-u_2, \;\; 
C_3=u_2-u_3, \dots, \; C_n=u_{n-1}-u_n, \;\;  \\
\textrm{or} \quad \;\; C_2=u_1-u_2, \;\; 
C_3=u_1-u_3, \dots, \; C_n=u_{1}-u_n. \nonumber
\end{eqnarray}
Integration constants
(\ref{sol_char}) have their counterparts in continuum formulation,
\begin{eqnarray}\label{sol_char_cont}
C_1 &=& \Phi \exp{[-u(x_1)]} \qquad  \textrm{or} \qquad
C_1 = \Phi \exp{\left(- \frac{1}{b-a}\int_a^b u(x) dx \right)}, 
\nonumber \\ 
  C(x) &=& \dd{u(x)}{x}\,\drm x  \qquad \textrm{or} \qquad
  C(x)=u(x_1)-u(x)
\end{eqnarray}
where $x_1$ is a fixed point in the domain 
$x\, \epsilon\, (a,b)$.
Hence, a functional that 
constitutes a solution of  Eq.~(\ref{eq_char}) in its continuum limit
may be written as
\begin{eqnarray}
F=F\left(\Phi\exp{(-u(x_1))},[u(x_1)-u(x)]\right).
\end{eqnarray}

\section{Finite and infinitesimal transformations.}

In this section we recall the 
classical symmetry method which can be used to analyse
 the partial differential equation (\ref{pde})
 and present its continuum extension for the
functional differential equation (\ref{fde}).
By ``symmetry transformation'' we understand such transformation
of variables which does not change the 
functional form of the considered
equation. This means that, for example Eq.~(\ref{fde})
in the old variables   $\Phi$, $t_1\dots t_n$, $\vu(\vx)$ and
the same equation written in  new, transformed 
variables $\vPhi$, $\tb_1\dots\tb_n$,  $\vub(\vx)$
\begin{equation}\label{fden}
F([\vub(\vx)],\tb_1,\dots, \tb_m,\vPhi,
\underset{1}{\vPhi},
\underset{2}{\vPhi},\dots,\underset{q}{\vPhi})=0;
\end{equation}
are equivalent. 
Note that $\vx$ is not transformed in the present approach
since it constitutes a continuous ``counting'' parameter,
such as in a summation for the classical counterpart.
Here, we consider only such 
transformations of variables which constitute Lie groups, i.e.\
they depend on a continuous parameter $\e$ and satisfy 
group properties, such as closure, associativity and containing
the unitary and inverse elements.

Table \ref{tab1} presents the comparison of  finite 
one-parameter Lie point transformation for the classical and
continuum formulation.
\noindent
\begin{table}
\caption{\label{tab1} Comparison of  
one-parameter Lie point transformation for the classical and
continuum formulation.}
\begin{tabular}{cll}
\hline
& classical formulation & continuum formulation \\
\hline
$i.$ &$\vPhi=\psi(\Phi,u_1,\dots, u_n,t_1,\dots, t_m,\e)$ &
$\vPhi =\psi(\Phi,[\vu(\vx)],t_1,\dots, t_m,\e)$\\
$ii.$ & $\ub_1=\phi_1(\Phi,u_1,\dots,u_n,t_1,\dots, t_m,\e)$ &
$\vub (\vx)=\vphi_{\vx}(\Phi,[\vu(\vx')],\vx,t_1,\dots, t_m,\e)$\\ 
 &$\; \vdots \qquad \vdots$ & $ \vx\,\epsilon\, {\bf G}$  \\
& $\ub_n=\phi_n(\Phi,u_1,\dots,u_n,t_1,\dots, t_m,\e)$ & \\
$iii.$ & $\tb_1=\phi_{t_1}(\Phi,u_1,\dots,u_n,t_1,\dots, t_m,\e)$ &
$\tb_1=\phi_{t_1}(\Phi,[\vu(\vx)],t_1,\dots, t_m,\e)$ \\
 &$\; \vdots \qquad \vdots$ & $\; \vdots \qquad \vdots$ \\
 & $\tb_m=\phi_{t_m}(\Phi,u_1,\dots,u_n,t_1,\dots, t_m,\e)$ &
$\tb_m=\phi_{t_m}(\Phi,[\vu(\vx)],t_1,\dots, t_m,\e)$ \\
\hline
\end{tabular}
\end{table}
As can be seen, the transformed variables 
$\vPhi$, $\vub (\vx)$, $\tb_1\dots\tb_m$ become
functionals in the continuum limit and
 depend on the infinite set of
independent variables $[\vu(\vx)]$.
It should also be noted that instead of the finite
set $\ub_1\dots\ub_n$,  
 in the continuum formulation we define
$\vub (\vx)=\vphi_{\vx}$, which is an explicit function
of the variable $\vx$, since $\vub$ defines a new variable at each
point $\vx$. 
This has important consequences in the
further considerations.

For the subsequent purpose of symmetry analysis
all variables of equation (\ref{pde}), i.e.\
the sets $t_1,\dots,t_m,u_1,\dots,u_n$, as well as
$\Phi$ and all its possible derivatives of any
order will be treated as independent
variables. Now, the
following, new differential operators are introduced:
\begin{eqnarray}\label{diffD}
\pcD{}{t_k} = 
\frac{\partial}{\partial t_k} + \Phi_{,t_k} 
\frac{\partial}{\partial \Phi } 
+ \sum_{j=1}^m \Phi_{,t_kt_j}\frac{\partial}{\partial \Phi_{,t_j} }
+ \sum_{j=1}^n\Phi_{,t_k u_j}\frac{\partial}{\partial \Phi_{,u_j} }
+ \cdot\cdot\cdot,\\
 k &=& 1,\dots,m \nonumber\\
 \frac{\mathcal{D}}{\mathcal{D} u_i} = 
\frac{\partial}{\partial u_i} + \Phi_{,u_i} 
\frac{\partial}{\partial \Phi } 
+ \sum_{j=1}^n\Phi_{,u_iu_j}\frac{\partial}{\partial \Phi_{,u_j} }
+  \sum_{j=1}^m \Phi_{,u_it_j}\frac{\partial}{\partial \Phi_{,t_j}} 
+ \cdot\cdot\cdot, \\
 i &=& 1,\dots,n. \nonumber
\end{eqnarray}
The partial derivatives e.g.\ of the form $\partial/\partial t_i$
will act only on terms which depends explicitly on $t_i$.
Within this  formulation
the derivatives of  $\Phi$ can be expressed as:
\begin{eqnarray}
\Phi_{,t_k} &=& \frac{\mathcal{D} \Phi }{\mathcal{D} t_k},  
\quad \Phi_{,u_i}  =
\frac{\mathcal{D} \Phi }{\mathcal{D} u_i}
\quad \Phi_{,u_i u_j}  =
\frac{\mathcal{D}  }{\mathcal{D} u_i}
\frac{\mathcal{D} \Phi }{\mathcal{D} u_j}, \\
\quad \Phi_{,u_it_k}  &=&
\frac{\mathcal{D}  }{\mathcal{D} u_i}
\frac{\mathcal{D} \Phi }{\mathcal{D} t_k}, 
\quad \Phi_{,t_jt_k}  =
\frac{\mathcal{D}  }{\mathcal{D} t_j}
\frac{\mathcal{D} \Phi }{\mathcal{D} t_k}. 
\end{eqnarray}
Analogous definitions will apply in the continuum limit
(\ref{fde}). The derivatives (\ref{diffD}) 
have the following counterparts
\begin{eqnarray}\label{deriv_c}
 \frac{\mathcal{D}}{\mathcal{D} t_k} &=& 
\frac{\partial}{\partial t_k} + \Phi_{,t_k} 
\frac{\partial}{\partial \Phi } +
\sum_{j=1}^m \Phi_{,t_kt_j}\frac{\partial}{\partial \Phi_{,t_j} }\nonumber \\
& + &
\sum_{\alpha=1}^{p}\int_{\bf G}  \drm\vx \,  
\Phi_{,t_ku_\alpha(\vx)}
\pdel{}{\Phi_{,u_\alpha(\vx)}} 
+ \cdot\cdot\cdot, \nonumber\\
\frac{\mathcal{D}}{\mathcal{D} u_\alpha(\vx) \dx} &=& 
\pdel{}{ u_\alpha(\vx)} +
\Phi_{,u_\alpha(\vx)} 
\frac{\partial}{\partial \Phi } 
+\sum_{j=1}^m 
\Phi_{,u_\alpha(\vx)\,t_j}
\frac{\partial}{\partial \Phi_{,t_j}} \\
&+&\sum_{\beta=1}^{p}
\int_{\bf G} \dx' \Phi_{,u_\alpha(\vx)u_\beta(\vx')}
\pdel{}{ \Phi_{,u_\beta({\bf x}')}} 
+ \cdot\cdot\cdot,\qquad
 \nonumber
\end{eqnarray}
(for  $\alpha=1,...,p, \quad  \vx\,\epsilon\, {\bf G}$)
and the derivatives of $\Phi$ in terms of the new differential operators
write:
\begin{eqnarray}\label{funder}
 \Phi_{,t_j}&=& \frac{\mathcal{D} \Phi }{\mathcal{D} t_j},  
\quad \Phi_{,u_\alpha(\vx)}  =
\frac{\mathcal{D} \Phi }{\mathcal{D} u_\alpha(\vx)  \dx}, \nonumber \\
\quad \Phi_{,u_\alpha(\vx) u_\beta(\vx')}  &=&
\frac{\mathcal{D}  }{\mathcal{D} u_\alpha(\vx) \dx}
\frac{\mathcal{D} \Phi }{\mathcal{D} u_\beta(\vx') \dx'},
\nonumber \\
\quad \Phi_{,  u_\alpha(\vx) t_j} & =&
\frac{\mathcal{D}  }{\mathcal{D} u_\alpha(\vx) \dx }
\frac{\mathcal{D} \Phi }{\mathcal{D} t_j},
\quad \Phi_{,t_jt_k}  =
\frac{\mathcal{D}  }{\mathcal{D} t_j}
\frac{\mathcal{D} \Phi }{\mathcal{D} t_k}.
\end{eqnarray}
Note that all derivatives in (\ref{funder}) are commutative.
Also it is important to distinguish between $\vx$ and $\vx'$
which denote different integration indices such as $i$ and
$j$ in two consecutive summations.
The quantities given by formulae ($i$), ($ii$), ($iii$) and
derivatives of $\vPhi$
can be written in a Taylor series expansion about $\e=0$.
Their infinitesimal forms, after neglecting terms of order 
$O(\e^2)$  are given in Table \ref{tab2}.
\noindent
\begin{table}
\caption{\label{tab2} Comparison of  
infinitesimal transformations for the classical and
continuum formulation.}
\begin{tabular}{@{}lll}
\hline
& classical formulation & continuum formulation \\
\hline
$iv.$ & $\vPhi=\Phi + \eta(\Phi,u_1,...,u_n,t_1,...,t_m)\, \e $ &
$\vPhi=\Phi + \eta(\Phi,[\vu(\vx)],t_1,...,t_m)\, \e $   \\
$v.$ & $\ub_1=u_1 + \xi_1(\Phi,u_1,...,u_n,t_1,...,t_m)\, \e  $ &
$\vub(\vx)=\vu(\vx)  $ \\
& $\, \vdots \qquad \vdots$&
$\qquad\, + 
\vxi_{\vx}(\Phi,[\vu(\vx')],\vx,t_1,...,t_m)\, \e$   \\
& $\ub_n=u_n + \xi_n(\Phi,u_1,...,u_n,t_1,...,t_m)\, \e  $ & \\
$vi.$ & $\tb_1=t_1 + \xi_{t_1}(\Phi,u_1,...,u_n,t_1,...,t_m)\, \e $ &
$\tb_1 = t_1 + \xi_{t_1}(\Phi,[\vu(\vx)],t_1,...,t_m)\, \e $ \\
& $\, \vdots \qquad \vdots$& $\, \vdots \qquad \vdots$   \\
 & $\tb_m=t_m + \xi_{t_m}(\Phi,u_1,...,u_n,t_1,...,t_m)\, \e $ &
$\tb_m = t_m + \xi_{t_m}(\Phi,[\vu(\vx)],t_1,...,t_m)\, \e $ \\
$vii.$ & $\vPhi_{,\tb_1}
= \Phi_{,t_1} + \zeta_{;t_1}(\Phi,u_1,...,u_n,t_1,...,t_m)\, \e  $&
$\vPhi_{,\tb_1} = \Phi_{,t_1} +
  \zeta_{;t_1}(\Phi,[\vu(\vx')],t_1,...,t_m)\, \e $ \\
& $\, \vdots \qquad \vdots$& $\, \vdots \qquad \vdots$   \\
& $\vPhi_{,\tb_m}
= \Phi_{,t_m} + \zeta_{;t_m}(\Phi,u_1,...,u_n,t_1,...,t_m)\, \e  $&
$\vPhi_{,\tb_m} = \Phi_{,t_m} +
  \zeta_{;t_m}(\Phi,[\vu(\vx')],t_1,...,t_m)\, \e $ \\
$viii.$ & $\vPhi_{,\ub_1}=
  \Phi_{,u_1} + \zeta_{;u_1}(\Phi,u_1,...,u_n,t_1,...,t_m)\, \e  $&
$\vPhi_{,\ub_\alpha(\vx)} = \Phi_{,u_\alpha(\vx)} $ \\
& $\, \vdots \qquad \vdots$& 
$+ \zeta_{;u_\alpha(\vx)}(\Phi,[\vu(\vx')],\vx,t_1,...,t_m)\, \e $  \\
& $\vPhi_{,\ub_n}=
  \Phi_{,u_n} + \zeta_{;u_n}(\Phi,u_1,...,u_n,t_1,...,t_m)\, \e  $&
$\alpha=1,...,p $ \\
& $\, \vdots \qquad \vdots$&   \\
$ix.$ &  $\vPhi_{,\ub_i \ub_j}=  
\Phi_{,u_i u_j}$ &
$\vPhi_{,\ub_\alpha(\vx) \ub_\beta(\vx')} 
= \Phi_{,u_\alpha(\vx)u_\beta(\vx')} $ \\
& $  \, \vdots \qquad 
 + \zeta_{;u_i u_j}(\Phi,u_1,...,u_n,t_1,...,t_m)\, \e  $
& $ + \zeta_{;u_\alpha(\vx)u_\beta(\vx')}
(\Phi,[\vu(\vx'')],\vx,\vx',t_1,...,t_m)\, \e$  \\
\hline
\end{tabular}
\end{table}
\noindent
A few notation particularities should be noted. Indices of
$\zeta$ are separated by a semicolon to distinguish it from derivatives.
Functional $\xi_\vx$ which denote infinitesimals corresponding
to $\vu$ is an explicit function of $\vx$. The same dependence
holds true for 
 infinitesimals corresponding to the functional
derivatives of $\Phi$, such as $\zeta_{;u(x)}$.
In these cases the index of the
set $[\vu(\vx)]$ has been given a different name such as
$[\vu(\vx')]$ to avoid confusion with the parameter $\vx$.
The remaining infinitesimals do not depend on $\vx$.

The key property of Lie group method is that 
the finite transformations, given by the formulae ($i$)--($iii$)
can be computed from their
 infinitesimal forms ($iv$)--($vi$) 
 (cf.\ Bluman and Kumei, 1989). 
According to the  Lie's first theorem,
the finite form of the transformation can be obtained 
by integrating the first order system of equations.
For the continuum formulation this system takes the form:
\begin{equation}
\dd{\vPhi}{\e}=\eta, \quad \dd{\tb_i}{\e}=\xi_{t_i},\; i=1,\dots,m, \quad
\dd{\ub_\alpha(\vx)}{\e}=\xi_{\alpha\vx}, \; \alpha=1,\dots,p, \;\;
\vx\, \epsilon\, {\bf G};
\end{equation}
where the latter equations should be integrated with the initial condition
\begin{equation}
\e=0:\; \vPhi=\Phi, \;\;  \tb_i=t_i, \;\; \ub_\alpha(\vx)=u_\alpha(\vx).
\end{equation}

Now, our aim is to find the infinitesimals forms $\eta$, $\xi_{t_i}$,
$\xi_{\alpha\vx}$.
To do this we should first express the infinitesimals $\zeta$
in terms of $\eta$, $\xi_{t_i}$, $\xi_{\alpha\vx}$
and independent variables
$t_1,\dots,t_m,[\vu(\vx)],\Phi$.
In the classical formulation
the following relation is used for this purpose
\begin{equation}
\frac{\mathcal{D} \psi }{\mathcal{D} u_i} =
\sum_{k=1}^n \frac{\mathcal{D} \phi_k}{\mathcal{D} u_i}
\pcDb{ \vPhi}{\ub_k} + \sum_{k=1}^{m}
 \frac{\mathcal{D} \phi_{t_k}}{\mathcal{D} u_i}
\pcDb{ \vPhi}{\tb_k} =
\sum_{k=1}^n \vPhi_{,\ub_k}  
\frac{\mathcal{D} \phi_{k}}{\mathcal{D} u_i}
+\sum_{k=1}^{m} \vPhi_{,\tb_k}
 \frac{\mathcal{D} \phi_{t_k}}{\mathcal{D} u_i};
\end{equation}
its continuum counterpart writes
\begin{equation}\label{psi_con}
\frac{\mathcal{D}\psi }{\mathcal{D} u_\alpha(\vx) \dx} = 
\sum_{\beta=1}^{p}\int_{\bf G}
 \vPhi_{,\ub_\beta(\vx')} 
 \frac{\mathcal{D}  
\phi_{\beta\vx'}(\Phi,[\vu(\vx)],\vx',t,\e) }
{\mathcal{D} u_\alpha(\vx) \dx}\dx'
+\sum_{k=1}^{m} \vPhi_{,\tb_k}
 \frac{\mathcal{D} \phi_{t_k}}{\mathcal{D}  u_\alpha(\vx) \dx}. 
\end{equation}
When the infinitesimal forms ($iv$)--($viii$), are introduced into
 equation (\ref{psi_con}) we obtain:
\begin{eqnarray} \label{phi_int}
\frac{\mathcal{D} (\Phi + \eta \e)  }{\mathcal{D} u_\alpha(\vx) \dx} &=& 
\sum_{\beta=1}^{p}\int_{\bf G}
 (\Phi_{,u_\beta(\vx')} + \zeta_{;u_\beta(\vx')}\e )
 \frac{\mathcal{D} (u_\beta(\vx') + \xi_{\beta\vx'}\e )  }
{\mathcal{D} u_\alpha(\vx) \dx}\dx' \nonumber  \\
&+& \sum_{k=1}^{m} (\Phi_{,t_k} + \zeta_{;t_k} \e)
 \frac{\mathcal{D} (t_k+ \xi_{t_k} \e)}{\mathcal{D}  
u_\alpha(\vx) \dx}. 
\end{eqnarray}
Equation (\ref{phi_int}) can be further split into two
equations, containing terms $O(1)$ and  $O(\e)$,
respectively. The first of the two gives the identity
\begin{eqnarray}
\frac{\mathcal{D} \Phi  }{\mathcal{D} u_\alpha(\vx) \dx} &=& 
\sum_{\beta=1}^{p}
\int_{\bf G} \Phi_{,u_\beta(\vx')}  
\frac{\mathcal{D} u_\beta(\vx') }{\mathcal{D} u_\alpha(\vx) \dx}\dx'
\nonumber \\
&=& \sum_{\beta=1}^{p} \delta_{\alpha\beta}
\int_{\bf G}  \Phi_{,u_\beta(\vx')}  \delta(\vx-\vx') \, \dx' =
 \Phi_{,u_\alpha(\vx)} 
\end{eqnarray}
where $\mathcal{D} u_\beta(\vx')/\mathcal{D} u_\alpha(\vx) \dx=
\delta_{\alpha\beta} \delta(\vx-\vx') $ has been used
where $\delta_{\alpha\beta}$ and 
$\delta(\vx-\vx')=\delta(x_1-x'_1)\cdot \cdots\, \delta(x_r-x_r')$
denote the Kronecker and Dirac delta respectively.
From $O(\e)$ we obtain a formula for the infinitesimals 
$\zeta_{;u_\alpha(\vx)}$ 
\begin{eqnarray} 
 \frac{\mathcal{D} \eta   }{\mathcal{D} u_\alpha(\vx) \dx} &=& 
\sum_{\beta=1}^{p}
\int_{\bf G} \left[ \Phi_{,u_\beta\vx'} 
 \frac{\mathcal{D} \xi_{\beta\vx'}   }
{\mathcal{D} u_\alpha(\vx) \dx}
+  \zeta_{;u_\beta(\vx')}
 \frac{\mathcal{D} u_\beta(\vx') }
{\mathcal{D} u_\alpha(\vx) \dx}\right] \dx' \nonumber \\
&+&  \sum_{k=1}^{m} \Phi_{,t_k} 
 \frac{\mathcal{D}  \xi_{t_k} }{\mathcal{D}  u_\alpha(\vx) \dx},
\end{eqnarray} 
hence,
\begin{equation}\label{zeta} 
 \zeta_{;u_\alpha(\vx)} = 
\frac{\mathcal{D} \eta   }{\mathcal{D} u_\alpha(\vx) \dx } -
\sum_{\beta=1}^{p}
\int_{\bf G}  \Phi_{,u_\beta(\vx')} 
 \frac{\mathcal{D} \xi_{\beta\vx'}   }
 {\mathcal{D} u_\alpha(\vx) \dx} \dx'
-  \sum_{k=1}^{m}  \Phi_{,t_k} 
 \frac{\mathcal{D}  \xi_{t_k} }{\mathcal{D}  u_\alpha(\vx) \dx}.
\end{equation}
By analogy, formula  for the infinitesimals
$\zeta_{;t_i}$ can be found
\begin{equation}\label{zetat} 
 \zeta_{;t_i} = 
\frac{\mathcal{D} \eta   }{\mathcal{D} t_i } -
\sum_{\beta=1}^{p}
\int_{\bf G}  \Phi_{,u_\beta(\vx')} 
 \frac{\mathcal{D} \xi_{\beta\vx'}   }
 {\mathcal{D} t_i} \dx'
-  \sum_{k=1}^{m}  \Phi_{,t_k} 
 \frac{\mathcal{D}  \xi_{t_k} }{\mathcal{D}  t_i}.
\end{equation}
The infinitesimals of higher orders will follow 
from the following, recursive formulae:
\begin{eqnarray}\label{zetap}
 \frac{\mathcal{D}\psi_{,u_\alpha(\vx^{(1)}),\dots,u_\beta(\vx^{(s-1)})} }
{\mathcal{D} u_\gamma(\vx^{(s)}) \dx^{(s)}} &=& 
\sum_{\delta=1}^{p}
\int_{\bf G}  \vPhi_{,u_\alpha(\vx^{(1)}),\dots,u_\beta(\vx^{(s-1)}),
u_\delta(\vx)}
 \frac{\mathcal{D} \phi_{\delta\vx} }
{\mathcal{D} u_\gamma(\vx^{(s)}) \dx^{(s)}}\,\dx \nonumber \\
&+& \sum_{k=1}^{m} \vPhi_{,u_\alpha(\vx^{(1)}),\dots,u_\beta(\vx^{(s-1)}),t_k}
 \frac{\mathcal{D} \phi_{t_k} }{\mathcal{D}  u_\gamma(\vx^{(s)}) \dx^{(s)}}. 
\end{eqnarray}
or
\begin{eqnarray}\label{zetapt}
\frac{\mathcal{D}\psi_{,u_\alpha(\vx^{(1)}),\dots,u_\beta(\vx^{(s-1)})} }
{\mathcal{D} t_j} &=& 
\sum_{\delta=1}^{p}
\int_{\bf G}  \vPhi_{,u_\alpha(\vx^{(1)}),\dots,u_\beta(\vx^{(s-1)}),
u_\delta(\vx)}
 \frac{\mathcal{D} \phi_{\delta\vx} }
{\mathcal{D} t_j}\,\dx \nonumber \\
&+& \sum_{k=1}^{m} \vPhi_{,u_\alpha(\vx^{(1)}),\dots,u_\beta(\vx^{(s-1)}),t_k}
 \frac{\mathcal{D} \phi_{t_k} }{\mathcal{D}  t_j}. 
\end{eqnarray}

\section{Generator $\Xrm$ and its prolongations.}

Once all the necessary infinitesimal forms are obtained
they can be  substituted into the  equations
(\ref{pde}) or (\ref{fde}) written in the
transformed variables.
In order to simplify notation we will 
assume that Eqs (\ref{pde}) and (\ref{fde})
only contain derivatives up to the second order.
The generalization of the following relations
to the case of higher order derivations is
straightforward. After  expansion in 
Taylor series about $\e=0$
in both, classical and continuum formulation the
expanded equation has the form:
\begin{equation}\label{texp}
F + \e \Xrm^{(2)}F + \frac{\e^2}{2}\left[\Xrm^{(2)}\right]^2 F
+ O(\e^3) = 0
\end{equation} 
where $\Xrm^{(2)}$ in  the classical formulation
is  given by the formula
\begin{eqnarray}\label{prolX}
\Xrm^{(2)} = \eta\pd{}{\Phi} + \sum_{j=1}^{m}\xi_{t_j}\pd{}{t_j}
+ \sum_{j=1}^{n}\xi_{j}\pd{}{u_j} + 
\sum_{j=1}^{m}\zeta_{;t_j}\pd{}{\Phi_{,t_j}}
+\sum_{j=1}^{n}\zeta_{;u_j}\pd{}{\Phi_{,u_j}} \nonumber\\
+\sum_{j=1}^{m}\sum_{k=1}^m\zeta_{;t_j t_k}\pd{}{\Phi_{,t_j t_k}}
+ \sum_{j=1}^{n}\sum_{k=1}^m \zeta_{;u_j t_k} \pd{}{\Phi_{,u_j t_k}}
+ \sum_{j=1}^{n}\sum_{k=1}^n \zeta_{;u_j u_k} \pd{}{\Phi_{,u_j u_k}} 
\end{eqnarray}
and is called the prolongation of the generator $\Xrm$
\begin{equation}\label{genX}
\Xrm = \eta\pd{}{\Phi} + \sum_{j=1}^{n}\xi_{t_j}\pd{}{t_j}
+ \sum_{j=1}^{m}\xi_{j}\pd{}{u_j}
\end{equation}
of the second order.
The corresponding formulae for the continuum limit write
\begin{eqnarray}\label{genXf}
\Xrm  = \eta\pd{}{\Phi} + \sum_{j=1}^{m}\xi_{t_j}\pd{}{t_j}
+ \sum_{\alpha=1}^{p}\int_{\bf G} 
\dx \, \xi_{\alpha \vx }\pdel{}{u_\alpha(\vx)} 
\end{eqnarray}
and
\begin{eqnarray}\label{xrmp}
\Xrm^{(2)}&=& \eta\pd{}{\Phi} + \sum_{j=1}^{m}\xi_{t_j}\pd{}{t_j}
+ \sum_{\alpha=1}^{p}\int_{\bf G} 
\dx \, \xi_{\alpha \vx }\pdel{}{u_\alpha(\vx)} + 
\sum_{j=1}^{m}\zeta_{;t_j}\pd{}{\Phi_{,t_j}}
\nonumber\\
&+&\sum_{\alpha=1}^{p}\int_{\bf G} \dx \,
\zeta_{;u_\alpha(\vx)}\pdel{}{\Phi_{,u_\alpha(\vx)}} 
+ \sum_{j=1}^{m}\sum_{k=1}^m\zeta_{;t_j t_k}\pd{}{\Phi_{,t_j t_k}}
\nonumber \\
&+& \sum_{\alpha=1}^{p}\sum_{k=1}^m \int_{\bf G} \dx \,
\zeta_{;u_\alpha(\vx) t_k} \pdel{}{\Phi_{,u_\alpha(\vx) t_k}}
\nonumber \\
&+& \sum_{\alpha=1}^{p}\sum_{\beta=1}^p \int_{\bf G}\int_{\bf G}\, \dx\, \dx'\,
\zeta_{;u_\alpha(\vx) u_\beta(\vx')} 
\pdel{}{\Phi_{,u_\alpha(\vx) u_\beta(\vx')}}.
\end{eqnarray}
The first term in Eq.~(\ref{texp}) equals zero, as
follows from the equations (\ref{pde}) or
(\ref{fde}). All the remaining terms $[\Xrm^{(2)}]^n$ in (\ref{texp}),
representing a successive application of $\Xrm^{(2)}$  will be zero if
the following relation holds
\begin{equation}
\Xrm^{(2)}F=0.
\end{equation}
In order to find the infinitesimal transformations  
we use the condition
\begin{equation}
\left. \left[ \Xrm^{(2)}F\right] \right|_{F=0}=0
\end{equation}
where, in the continuum formulation,
the prolongation $\Xrm^{(2)}$ is expressed
by the formula (\ref{xrmp}) and the forms of infinitesimals
$\zeta$ are found from relations (\ref{zeta})--(\ref{zetapt}). 
The resulting condition constitutes an overdetermined
 system of linear differential equations.
In the continuum limit we obtain a set of functional differential 
equations.
This system can be further
solved for the infinitesimals $\eta$, $\xi_{t_j}$ and
$\xi_{\alpha\vx}$.

\section{Invariant solutions}
If the functional differential equation  (\ref{fde})
admits a symmetry given by the generator
(\ref{genXf}), then  a solution 
$\Phi=\Theta(t_1,\dots,t_m,[u(x)])$  of this equation
 is called an invariant solution if
it  satisfies the relation
\begin{equation}\label{inv_sol}
\Xrm \left[\Phi - \Theta\left( t_1,\dots,t_m,[u(x)] \right) \right] = 0.
\end{equation}
After employing (\ref{genXf}) and expanding the
derivatives, from (\ref{inv_sol}) the following,
hyperbolic functional equation is obtained
\begin{equation}\label{hyper_Theta}
\sum_{i=1}^m \xi_{t_i}\pd{\Theta}{t_i} +
\sum_{\alpha=1}^p \int_{\bf G} 
\xi_{\alpha \vx}\pdel{\Theta}{u_\alpha(\vx)}\, \drm \vx
= \eta
\end{equation}
This equation can be solved by the method of characteristics.
The corresponding system of equations writes
\begin{equation}\label{Fun_char_sys}
\frac{\drm t_1}{ \xi_{t_1}}=\dots=\frac{\drm t_m}{ \xi_{t_m}}
=\frac{\drm \Phi}{ \eta }=\frac{\delta{u_\alpha(\vx)}}{\xi_{\alpha \vx}}
\qquad \textrm{for each} \qquad \vx \; \epsilon \; {\bf G}
\;\; \textrm{and} \;\; \alpha=1,\dots,p.
\end{equation}
Above, $\Theta$ has been replaced by $\Phi$.
Note that the last term in fact corresponds to an
infinite set of equations for each $\alpha$ and each point
in ${\bf G}$.
The infinite set of constants, which is  a
solution of the above system, can be employed as new
variables in Eq.~(\ref{fde}). As in the considered case
one of them, say $C_1$, will be a dependent variable and the
rest will constitute a set of independent variables,
the following relation holds 
\begin{eqnarray}\label{rel_cons}
C_1=H(C_2,\dots,C_m,[C_1(\vx)],\dots,[C_p(\vx)]).
\end{eqnarray}
 After the process of solving 
characteristic system for a partial differential equation
with a finite set of variables, the number of
independent variables is reduced by one.
In the case of functional differential equations,
in formula (\ref{rel_cons}) 
one point of the considered domain will be excluded
from a set $[C_i(\vx)]$, 
 $\vx \; \epsilon \; {\bf G} / \{x_{j1}\}$.

\section{Example}
For the sake of clarity we will consider, as an example,
the continuum limit
of a heat equation in infinite many dimension. 
In a classical formulation we take $m=1$.
Hence, the equation for a function of
 $n+1$ variables $\Phi=\Phi(u_1,...,u_n,t)$ writes
\begin{equation}\label{h_d}
\frac{\partial \Phi}{\partial t}=
\sum_{i=1}^n \frac{\partial^2 \Phi}{\partial u_i^2},
\end{equation}
while in the continuum limit we  consider
\begin{equation}\label{h_c}
\frac{\partial \Phi}{\partial t}=
\int_a^b \frac{\partial}{\partial u(x) \drm x}
 \frac{\partial \Phi}{\partial u(x) \drm x} \, \drm x
\qquad \textrm{or} \qquad
\Phi_{,t}=\int_a^b\Phi_{,u(x)u(x)}\drm  x
\end{equation}
where $\Phi=\Phi([u(x)],t)$.
The generator $\Xrm$ of Eq.~(\ref{h_c}) 
(cf.\ Eq.~(\ref{genX})) has the form
\begin{equation}
\Xrm = \eta \pd{}{\Phi} + \xi_t\pd{}{t} + 
\int_a^b \drm x \, \xi_{x} \pdelu{}{}.
\end{equation}
The second 
prolongation $\Xrm^{(2)}$ of $\Xrm$ necessary for (\ref{h_c})
(cf.\ Eq.~(\ref{prolX})) is defined as
\begin{eqnarray}\label{ex_X2}
\Xrm^{(2)} &=&  \eta \pd{}{\Phi} + \xi_t\pd{}{t} +
\int_a^b \drm x \, \xi_{x} \pdelu{}{}+
 \zeta_{;t}\pd{}{\Phi_{,t}} \nonumber \\
&+&
\int_a^b \drm x\, \zeta_{;u(x)u(x)} \,  \pdel{}{\Phi_{,u(x)u(x)}},
\end{eqnarray}
where any unneeded $\zeta$ has been omitted.
Applying (\ref{ex_X2}) to (\ref{h_c}) the first three
terms of  (\ref{h_c}) have no effect, while
the fourth one acts on $\Phi_{,t}$ to lead to $ \zeta_{;t}$.
The last term in  (\ref{ex_X2}), acting on (\ref{h_c}) can be written as
\begin{eqnarray}
&&\int_a^b \drm x \zeta_{;u(x)u(x)} \frac{\delta}{\delta \Phi_{,u(x)u(x)}}
\int_a^b  \Phi_{,u(x')u(x')}  \drm x' = \\&&
\int_a^b \drm x \zeta_{;u(x)u(x)} \int_a^b \delta(x-x')\drm x' =
 \int_a^b\zeta_{;u(x)u(x)} \, \drm x. \nonumber
\end{eqnarray} 
As a result we obtain
\begin{equation}
 \zeta_{;t}- \int_a^b\zeta_{;u(x)u(x)} \, \drm x =0.
\end{equation}
Into the above equation we  substitute the
infinitesimals $\zeta_{;t}$, $\zeta_{;u(x) u(x)}$,
found from Eqs (\ref{zeta})--(\ref{zetap}). Their 
forms, without derivation, are given below
\begin{eqnarray} 
 \zeta_{;t} &=& 
\frac{\mathcal{D} \eta   }{\mathcal{D} t } -
\int_a^b  \Phi_{,u(x)} 
 \frac{\mathcal{D} \xi_{x}   }{\mathcal{D} t } \drm x
-   \Phi_{,t} 
 \frac{\mathcal{D}  \xi_t }{\mathcal{D}  t }, \nonumber \\
 \zeta_{;u(x) u(x)} &=& 
\frac{\mathcal{D}        }{\mathcal{D} u(x) \drm x}
\frac{\mathcal{D} \eta   }{\mathcal{D} u(x ) \drm x } -
\frac{\mathcal{D}        }{\mathcal{D} u(x) \drm x}
\int_a^b  \Phi_{,u(x')} 
 \frac{\mathcal{D} \xi_{x'}   }{\mathcal{D} u(x) \drm x } \drm x' \nonumber  \\
&-& \int_a^b  \Phi_{,u(x) u(x')} 
 \frac{\mathcal{D} \xi_{x'}   }{\mathcal{D} u(x) \drm x} \drm x' 
-  \Phi_{,u(x) t}  \frac{\mathcal{D} \xi_{t}   }{\mathcal{D} u(x) \drm x}
 \nonumber  \\
&-& \frac{\mathcal{D}        }{\mathcal{D} u(x) \drm x}
\left( \Phi_{,t} 
 \frac{\mathcal{D}  \xi_t }{\mathcal{D}  u(x) \drm x} \right). \nonumber
\end{eqnarray}
If the definitions of differential operators  
(\ref{deriv_c}) are used in the above equations
one obtains
\begin{eqnarray} 
 \zeta_{;t} &=& \pd{\eta}{t}+\Phi_{,t} \pd{\eta}{\Phi}
-\Phi_{,t}\pd{\xi_t}{t}-\Phi^2_{,t}\pd{\xi_t}{\Phi} 
\nonumber\\
&-&\int_a^b\left( 
\Phi_{,u(x)}\pd{\xi_{x}}{t}+\Phi_{,u(x)}\Phi_{,t}\pd{\xi_{x}}{\Phi} 
\right)\, \drm x \quad \textrm{and} \nonumber\\
 \zeta_{;u(x) u(x)} &=& \pdel{^2\eta}{u(x)^2} +
2\Phi_{,u(x)}\pdelu{}{}\pd{\eta}{\Phi} + \Phi^2_{,u(x)}
\pd{^2\eta}{\Phi^2} + \pd{\eta}{\Phi}\Phi_{,u(x)u(x)}\nonumber  \\
& -& \int_a^b \Phi_{,u(x)u(x')} \left( 
\pdelu{\xi_{x'}}{} + \Phi_{,u(x)}\pd{\xi_{x'}}{\Phi}\right)\,\drm x'
 \\
& -& \int_a^b \Phi_{,u(x')} \left(
\pdel{^2\xi_{x'}}{u(x)^2} + 2 \Phi_{,u(x)}\pdelu{}{} 
\pd{\xi_{x'}}{\Phi} + \Phi^2_{,u(x)}\pd{^2\xi_{x'}}{\Phi^2}
+ \Phi_{,u(x)u(x)} \pd{\xi_{x'}}{\Phi}
\right)\,\drm x'\nonumber \\
& -& \Phi_{,u(x)t}\left( \pdel{\xi_t}{u(x)} + \Phi_{,u(x)}\pd{\xi_t}{\Phi}
\right) \nonumber \\ 
 & -& \Phi_{,t}\left(\pdel{^2\xi_t}{u(x)^2} +
2\Phi_{,u(x)}\pdelu{}{}\pd{\xi_t}{\Phi}
+ \Phi^2_{,u(x)}\pd{^2\xi_t}{\Phi^2} 
+ \Phi_{,u(x)u(x)}\pd{\xi_t}{\Phi}
 \right) 
\nonumber \\
 &-& \int_a^b \Phi_{,u(x)u(x')} \left(\pdelu{\xi_{x'}}{}
+ \Phi_{,u(x)}\pd{\xi_{x'}}{\Phi}
\right) \drm x'
- \Phi_{,u(x)t}\left(\pdelu{\xi_t}{} + \Phi_{,u(x)}\pd{\xi_t}{\Phi}
\right)\nonumber.
\end{eqnarray}
Next, we use  Eq.~$(\ref{h_c})$ in order to 
substitute for the derivative $\Phi_{,t}$
\begin{eqnarray}
\Phi_{,t}=\int_a^b\Phi_{,u(x)u(x)}\,\drm x.
\end{eqnarray}
The resulting equation has the form
\begin{eqnarray}\label{inteq}
 \pd{\eta}{t}&-&
\int_a^b\pdel{^2\eta}{u(x)^2}\,\drm x 
%
-\int_a^b \Phi_{,u(x)}\left(\pd{\xi_{x}}{t}
+ 2 \pd{^2\eta}{\Phi\partial\, u(x)\,\drm x}
-\int_a^b\pdel{^2\xi_{x}}{u(x')^2}\,\drm x'
 \right)\,\drm x   \nonumber \\
&+&\int_a^b\Phi_{,u(x)u(x)} \left(
\int_a^b \pdel{^2\xi_t}{u(x')^2}\,\drm x'
-\pd{\xi_t}{t} \right)\,\drm x
+2 \int_a^b\int_a^b\Phi_{,u(x)u(x')}\pdel{\xi_{x'}}{u(x)}\,\drm x\,\drm x' 
  \nonumber  \\
&+&2\int_a^b\int_a^b\Phi_{,u(x)}\Phi_{,u(x')}
\pd{^2\xi_{x'}}{\Phi\partial u(x)\drm x}\,\drm x\,\drm x'
-\int_a^b\Phi_{,u(x)}^2\pd{^2\eta}{\Phi^2}\,\drm x    \\
&+&2\int_a^b\int_a^b\Phi_{,u(x)u(x')}\Phi_{,u(x)}
\pd{\xi_{x'}}{\Phi}\,\drm x\,\drm x'
+2 \int_a^b\int_a^b  \Phi_{,u(x')u(x')}  \Phi_{,u(x)} 
\pd{^2\xi_t}{\Phi\partial u(x)\drm x}\,\drm x\,\drm x'
 \nonumber \\
&+&\int_a^b\int_a^b \Phi^2_{,u(x)}\Phi_{,u(x')} \pd{^2\xi_{x'}}{\Phi^2}
\,\drm x\,\drm x' 
+ \int_a^b\int_a^b  \Phi_{,u(x')u(x')}  \Phi_{,u(x)}^2 
\pd{^2\xi_t}{\Phi^2}\,\drm x\,\drm x' \nonumber \\
&+&2\int_a^b \Phi_{,u(x)t} \pdel{\xi_t}{u(x)}\,\drm x 
+2\int_a^b \Phi_{,u(x)t} \Phi_{,u(x)}\pd{\xi_t}{\Phi}\,\drm x = 0.
 \nonumber
\end{eqnarray}
%
As the terms $\eta$, $\xi_t$, $\xi_{x}$ do not depend on
the derivatives of $\Phi$, from Eq.~(\ref{inteq}) the following
system of differential equations can be obtained, where on
the left hand side the coefficient function is written:
\begin{eqnarray}
 \Phi_{,u(x)t} \Phi_{,u(x)}: & \pd{\xi_t}{\Phi} = 0, \label{E1} \\
 \Phi_{,u(x)t}:  &\pdel{\xi_t}{u(x)}= 0, \quad \textrm{for each}
\quad x \, \epsilon\, (a,b), \label{E2} \\
 \Phi_{,u(x)u(x')} \Phi_{,u(x)}: 
&\pd{\xi_{x}}{\Phi}= 0,  \quad \textrm{for each}  
 \quad x \, \epsilon\, (a,b),  \label{E3} \\
 \Phi^2_{,u(x)}: &\pd{^2\eta}{\Phi^2}=0, \label{E4} \\
 \Phi_{,u(x)u(x)}: &\int_a^b \pdel{^2\xi_t}{u(x')^2}\,\drm x'
-\pd{\xi_t}{t} + 2 \pdel{\xi_{x}}{u(x)} = 0, \label{E5} \\ 
& \quad \textrm{for each} \quad x \, \epsilon\, (a,b),
 \nonumber \\ 
 \Phi_{,u(x)u(x')}: &\pdel{\xi_{x}}{u(x')}  +
 \pdel{\xi_{x'}}{u(x)} = 0, \label{E6} \\
&  \quad \textrm{for each}
\quad x,x' \, \epsilon\, (a,b),  \; \textrm{and} \; x \ne x',
 \nonumber \\
 \Phi_{,u(x)}: &\pd{\xi_{x}}{t}
+ 2 \pd{^2\eta}{\Phi\partial\, u(x)\,\drm x}
-\int_a^b\pdel{^2\xi_{x}}{u(x')^2}\,\drm x'  = 0, \label{E7} \\
& \; \textrm{for each} \;
 x \, \epsilon\, (a,b) \nonumber \\
 1:& \pd{\eta}{t}- \int_a^b\pdel{^2\eta}{u(x)^2}\,\drm x =0. \label{E8} 
\end{eqnarray}
Both equations (\ref{E5}) and (\ref{E6}) follow from 
the second line of Eq.~(\ref{inteq}). To see that this is true we
will differentiate Eq.~(\ref{inteq}) with respect to
by $\delta/\delta \Phi_{,u(x'')u(x''') }$.
Since $\Phi_{,u(x)u(x') }$ is an explicit function of two variables
it follows that
\begin{equation}
\pdel{ \Phi_{,u(x)u(x') }}{ \Phi_{,u(x'')u(x''') }}=
\delta(x-x'',x'-x''')=\delta(x-x'')\,\delta(x'-x''').
\end{equation}
On the other hand, $\Phi_{,u(x)u(x)}$ is a function
of variable $x$ only, say $\Phi_{,u(x)u(x)}=g(x)$, hence
\begin{equation}\label{sol}
\pdel{ \Phi_{,u(x)u(x) }}{ \Phi_{,u(x'')u(x'') }}=
\pdel{g(x)}{g(x'')}= \delta(x-x'').
\end{equation}
The functional derivative of $\Phi_{,u(x)u(x) }$ 
with respect to $\Phi_{,u(x'')u(x''') }$ is zero 
if $x''\ne x'''$. Finally, 
after differentiation of (\ref{inteq}) with respect to
$\Phi_{,u(x'')u(x''') }$, for $x''=x'''$ we obtain 
\begin{eqnarray} \label{E5d}
&\int_a^b\delta(x-x'')\, \drm x \left(
\int_a^b \pdel{^2\xi_t}{u(x')^2}\,\drm x'
-\pd{\xi_t}{t} \right)\\
&+2 \int_a^b\int_a^b \delta(x-x'')\delta(x'-x'')
\pdel{\xi_{x'}}{u(x)}\,\drm x\,\drm x'=0 \nonumber 
\end{eqnarray}
whereas in case  when $x'' \ne x'''$
we have 
\begin{eqnarray} \label{E6d}
2 \int_a^b\int_a^b \delta(x-x'')\delta(x'-x''')
\pdel{\xi_{x'}}{u(x)}
+ \delta(x'-x'')\delta(x-x''')
\pdel{\xi_{x}}{u(x')}
\,\drm x\,\drm x'=0. \nonumber 
\end{eqnarray}
Note that the presence of two terms inside the integral
is a consequence of an equality
 $\Phi_{,u(x)u(x') }=\Phi_{,u(x')u(x)}$. 
From the above equations formulae 
(\ref{E5}) and (\ref{E6}) are clearly obtained. 

The above system of equations is now solved in order to find 
the form of infinitesimals $\eta$, $\xi_t$ and $\xi_{x}$.
From relations (\ref{E1})--(\ref{E4}) we obtain:
\begin{eqnarray}
\xi_t &=& \xi_t(t), \\
\xi_{x} &=& \xi_{x}([u(x')],x,t), \\
\eta &=&  f_1([u(x')],t)\,\Phi +  f_2([u(x')],t), \label{f_eta}
\end{eqnarray}
Next, we consider formula (\ref{E5}).
We recall that $\xi_x$ is an explicit function of $x$, hence,
 the general solution of  (\ref{E5}) has the form
\begin{eqnarray}\label{xi_E5}
\xi_{x} = \frac{1}{2}\frac{d\xi_t}{dt}\,\int_a^b u(x') \, \drm x' + 
H(x,t,[u(x')]) + f_3(x,t)
\end{eqnarray}
where $H(x,t,[u(x')])$ is a functional depending explicitly on $x$,
such that its functional derivative 
$\delta H(x,t,[u(x')]) / \delta u(x) = 0$.
Without an
explicit dependence of $\xi_x$
on $x$ the second term on the RHS would
disappear as we would obtain $H(t,[u(x')])=0$.
Now, the form (\ref{xi_E5}) is introduced into Eq.~(\ref{E6})
to give
\begin{eqnarray}\label{E6H}
\frac{1}{2}\frac{d\xi_t}{dt} + \pdel{H(x,t,[u(x'')])}{u(x')}
= - \frac{1}{2}\frac{d\xi_t}{dt} - \pdel{H(x',t,[u(x'')])}{u(x)},
\quad \textrm{for} \quad x \ne x'.
\end{eqnarray}
Below, we prove that the functional derivative
of $H(x,t,[u(x')])$ with respect to $u(x')$ 
is a function of three variables $G(x,x',t)$.
First, we differentiate Eq.\ (\ref{E6H}) with respect to $u(x')$
and use the property $\delta H(x',t,[u(x'')]) / \delta u(x') = 0$
to obtain
\begin{eqnarray}
 \pdel{^2 H(x,t,[u(x'')])}{u(x')^2}=0.
\end{eqnarray}
Let us now differentiate Eq.\ (\ref{E6H}) with respect to
$u(x_1)$, such that $x_1 \ne x \ne x'$. We obtain
\begin{eqnarray}\label{derx1x}
 \pdel{^2 H(x,t,[u(x'')])}{u(x') \,\delta u(x_1)}=
-  \pdel{^2 H(x',t,[u(x'')])}{u(x) \,\delta u(x_1)}.
\end{eqnarray}
Now, if we note that
Eq.\ (\ref{E6H}) can be written also for pairs of variables
$x_1$, $x$ and $x_1$, $x'$, the LHS of the 
above formula equals
\begin{eqnarray}\label{derx}
\pdel{}{u(x')} \left(
\pdel{ H(x,t,[u(x'')])}{ u(x_1)} \right)=
-  \pdel{}{u(x')}\left(\pdel{ H(x_1,t,[u(x'')])}{ u(x)}\right)
\end{eqnarray} 
and the RHS is
\begin{eqnarray}\label{derx1}
-\pdel{}{u(x)}\left(
 \pdel{ H(x',t,[u(x'')])}{ u(x_1)} \right)=
  \pdel{}{u(x)}\left(
  \pdel{ H(x_1,t,[u(x'')])}{ u(x')}\right)
\end{eqnarray} 
which is an expression with opposite sign to that
in (\ref{derx}).
It follows that the second order functional derivative
of $H$ is always zero, hence
\begin{eqnarray}
\pdel{ H(x,t,[u(x'')])}{ u(x')}=G(x,x',t).
\end{eqnarray} 
From Eq.\ (\ref{xi_E5}) we obtain the relation
 $G(x,x',t)=-G(x',x,t)-\drm \xi_t / \drm t$,
additionaly $G(x,x,t)=0$, as the first derivative
of $H(x,t,[u(x'')])$ with respect to $u(x)$ is zero.
Now, the formula (\ref{xi_E5}) for the infinitesimal 
$\xi_x$ writes
\begin{eqnarray}
\xi_{x} = \frac{1}{2}\dd{\xi_t}{t}\,\int_a^b u(x') \, \drm x' + 
\int_a^b  G(x,x',t) u(x') \, \drm x' + f_3(x,t)
\end{eqnarray}
The above form
can be rearranged as follows
\begin{eqnarray}
\xi_{x} = 
\int_a^b  C(x,x',t) u(x') \, \drm x' + f_3(x,t)
\end{eqnarray}
where
\begin{eqnarray}\label{fC}
C(x,x',t)&=&-C(x',x,t). \qquad \textrm{for} \qquad x\ne x' \\
C(x,x,t)&=& \frac{1}{2}\dd{\xi_t}{t}, 
\end{eqnarray}

This result, together with (\ref{f_eta}) can be
 further substituted into (\ref{E7}) to give
\begin{eqnarray}\label{df1}
\pdel{ f_1([u(x')],t)}{u(x)} 
= -\frac{1}{2}    
\int_a^b \pd{C(x,x',t)}{t} u(x') \drm x'
-\frac{1}{2} \pd{f_3}{t}.
\end{eqnarray}
After integration we obtain
\begin{eqnarray}\label{f1}
& f_1([u(x')],t)
= -\frac{1}{4}  \int_a^b \int_a^b \pd{C(x,x',t)}{t} 
u(x') u(x) \drm x' \drm x \\
&-\frac{1}{2}\int_a^b \pd{f_3}{t} u(x') \, \drm x' + f_4(t). \nonumber
\end{eqnarray}
If we recall the differentiation rules presented in Section 
\ref{SecNotation} we find that the functional derivative
of the first RHS term in the above equation is a sum
of two integrals (cf.\ Eq.~(\ref{sec_der})) and
is equal to the corresponding term in Eq.~(\ref{df1})
only if for $x \ne x'$, 
$\partial C(x,x',t)/\partial t=\partial C(x',x,t)/\partial t$.
It further follows from (\ref{fC}) 
that the only possibility is 
$\partial C(x,x',t)/\partial t=0$ for $x \ne x'$.
Finally, the form (\ref{f_eta}) is substituted
into equation (\ref{E8}) to  give
\begin{eqnarray}
&\Phi \left(-\frac{1}{8}\frac{d^3\xi_t}{dt^3}
\int_a^b \int_a^b c(x,x') u(x) u(x') \,\drm x \drm x' \right. \nonumber \\
&\left.
-\frac{1}{2}\int_a^b \pd{^2f_3}{t^2} u(x)\,\drm x+\frac{df_4}{dt}
+ \frac{1}{4} \frac{d^2\xi_t}{dt^2}(b-a)  \right) \nonumber \\
&+ \pd{f_2}{t}-\int_a^b \pdel{^2 f_2}{u(x)^2}\,\drm x = 0
\end{eqnarray}
where $c(x,x)=1$ and $c(x,x') = 0$ for $x \ne x'$.
As $\xi_t$, $f_3$ and $f_4$  neither depend on
$\Phi$ nor on $u(x)$, from the above equation it follows that
\begin{eqnarray}
\frac{d^3\xi_t}{dt^3} &=& 0 \\
 \pd{^2f_3(x,t)}{t^2} &=& 0 \\
\frac{df_4}{dt} + \frac{(b-a)}{4} \frac{d^2\xi_t}{dt^2} &=& 0 \\
\pd{f_2}{t}-\int_a^b \pdel{^2 f_2}{u(x)^2}\,\drm x &=& 0.
\end{eqnarray}
Hence, the final solution of the system (\ref{E1})--(\ref{E8})
writes
\begin{eqnarray}
\xi_t&=&a_1 t^2 + a_2 t + a_3 \label{I1}\\
\xi_{x}&=&  \int_a^b C(x,x',t) u(x')\,\drm x'
+ a_4(x) t + a_5(x) \label{I2} \\
\eta &=& -\left(\frac{1}{4} a_1 \int_a^b \int_a^b c(x,x') u(x) u(x') 
\,\drm x\, \drm x'+ \right. \nonumber \\
&&\left(
\frac{1}{2} \int_a^b a_4(x) u(x)\,\drm x +
\frac{(b-a)}{2} a_1 t + a_6
\right) \Phi  \label{I3} 
+ f_2([u(x)],t)
\end{eqnarray} 
where $C(x,x,t)=  \frac{1}{2} \left(2a_1 t + a_2 \right)$,
$C(x,x',t)=C(x,x')=-C(x',x)$, 
and $f_2$ is a solution of equation (\ref{h_c}).

The infinitesimals (\ref{I1})--(\ref{I3}) may directly
be compared to those of the classical heat equation
(\ref{h_d}) (cf.\ Ibragimov 1995)
\begin{eqnarray}
\xi_t\!\!&=&\!\!a_1 t^2 + a_2 t + a_3 \label{I1d}\\
\xi_{i}\!\!&=&\!\! \frac{1}{2} \left(2a_1 t + a_2 \right)u_i + 
\sum_{j=1}^n C_{ij} u_j
+ a_{i4} t + a_{i5}, \quad \textrm{i=1,\dots,n} \label{I2d} \\
\eta\!\! &=&\!\! -\left(\frac{1}{4} a_1 \sum_{i=2}^n u_i^2 +
\frac{1}{2} \sum_{i=1}^n a_{i4} u_i +
\frac{1}{2}n a_1 t + a_6
\right) \Phi 
 + f_2(u_1,...,u_n,t) \label{I3d} 
\end{eqnarray}
where $C_{ij}=-C_{ji}$ for $i \ne j$ and $C_{ii}=0$.
As it is seen, the form of infinitesimal $\xi_t$, Eq.\ (\ref{I1d}), 
is the same as in the continuum limit. In formula (\ref{I2d})
the two first terms on the RHS can be rearranged as
$1/2(2 a_1 t + a_2)\sum_{i=1}^n C'_{ij} u_j$ with
 $C'_{ij}=-C'_{ji}$ for $i \ne j$ and $C'_{ii}=1$.
Then, in the continuum limit we obtain the first RHS term of
  (\ref{I2}). The constants $a_{i4}$ and $a_{i5}$ become
in (\ref{I2})  functions of a continuous parameter $x$.
The double integral in  Eq.\ (\ref{I3}) is a continuum 
counterpart of the first RHS term in (\ref{I3d}).
We note that
the second order derivatives of these terms,
present in Eq.~(\ref{E8}) and its classical counterpart, 
are the same in both formulations  
\begin{eqnarray}
\pd{f_1}{u_k} &=& -\frac{1}{2} a_1  u_k 
               -\frac{1}{2} a_{k4},   \nonumber  \\
\pd{^2 f_1}{u_k^2}& =& -\frac{1}{2} a_1,  \nonumber  \\
\pdel{f_1}{u(x)} &=&  -\frac{1}{4} a_1 \int_a^b c(x,x')\,\drm x'
 -\frac{1}{4} a_1 \int_a^b c(x',x)\,\drm x'
 -\frac{1}{2} a_{4}(x),  \nonumber  \\
\pdel{^2 f_1}{u(x)^2}& =&  -\frac{1}{2} a_1. \nonumber
\end{eqnarray} 
The second RHS term in Eq.~(\ref{I3d}) becomes an integral
in  Eq.\ (\ref{I3}). In the
third RHS term, instead of the number of variables $n$,
in the continuum formulation (\ref{I3}) we obtain the 
length of the considered integration domain which equals $(b-a)$.

From Eq.~(\ref{I1})-(\ref{I3}) we can distinguish the following 
symmetry groups, admitted by the considered Eq.~(\ref{h_c})
\begin{eqnarray}
\Xrm_t &=& \pd{}{t}, \label{S1} \\
\Xrm_{u(x)}& =& \pdel{}{u(x)}, 
\hskip 2.15cm \qquad\qquad \textrm{for each}
\quad x\,\epsilon\,(a,b), \label{S2} \\
\Xrm_{u(x)u(x')}& =&  u(x') \pdel{}{u(x)} -  u(x) \pdel{}{u(x')},
 \label{S3}\\
&&\quad \textrm{for each}\quad x,x'\,\epsilon\,(a,b) 
\quad \textrm{and} \quad
x \ne x', \nonumber \\
\Xrm_{u(x)\Phi} &=& 2t\pdel{}{u(x)} - u(x)\Phi\pd{}{\Phi}, 
\hskip 0.35cm
\qquad \textrm{for each}\quad x\,\epsilon\,(a,b),  \label{S4}\\
\Xrm_{tu(x)} &=& 2t\pd{}{t} + u(x)\pdel{}{u(x)}, 
\hskip 0.8cm
\qquad \textrm{for each}\quad x\,\epsilon\,(a,b), \label{S5}\\
\Xrm_{\Phi} &=& \Phi\pd{}{\Phi},  \label{S6}\\ 
\Xrm_{u(x)t\Phi} &=& t^2\pd{}{t} + t u(x)\pdel{}{u(x)} -
\frac{1}{4} \int_a^b \int_a^b c(x,x') u(x) u(x')\,\drm x\, \drm x' 
\;\Phi\pd{}{\Phi} \nonumber\\
&& - \frac{(b-a)}{2}t\Phi\pd{}{\Phi}, \qquad
\textrm{for each}\quad x\,\epsilon\,(a,b). \label{S7}
\end{eqnarray}

We will investigate the invariant solutions of Eq.\ (\ref{h_c})
under the combined symmetries (\ref{S2}), (\ref{S4})
for each $x\, \epsilon\, (a,b)$  and (\ref{S6}). 
The invariant solution of Eq.~(\ref{h_c}) satisfies
the following, hyperbolic equation (cf.\ formula (\ref{hyper_Theta}))
\begin{eqnarray}\label{inv_hc}
\int_a^b \left(a_4(x)t + a_5(x)\right) \pdel{\Phi}{u(x)} \drm x = 
-a_6 \Phi - \frac{1}{2} \Phi \int_a^b a_4(x')u(x')\drm x'
\end{eqnarray} 
the characteristic system writes
\begin{eqnarray}
\frac{\drm t}{0}=
\frac{\drm \Phi}{ -a_6 \Phi - \frac{1}{2} \Phi \int_a^b a_4(x')u(x')\drm x'}
= \frac{\delta u(x)}{a_4(x)t+a_5(x)} \qquad 
\end{eqnarray}
We will solve the following equations
\begin{eqnarray}
\drm t = 0, \\
 \frac{\delta u(x)}{a_4(x)t+a_5(x)}= 
\frac{\delta u(x_1)}{a_4(x_1)t+a_5(x_1)}, 
 \qquad \textrm{for fixed} \; x_1  \\
\frac{\drm \Phi}{ \Phi }
=- \frac{\int_a^b  \delta u(x) \left( 2 a_6 +  
\int_a^b a_4(x')u(x')\drm x' \right)  \drm x}
{2 \int_a^b \left( a_4(x)t+a_5(x)\right) \drm x} 
\end{eqnarray}
which give a set of integration constants
\begin{eqnarray}
 t = \tau, \\
 C(x) =  \frac{ u(x)}{a_4(x)t+a_5(x)}-
\frac{ u(x_1)}{a_4(x_1)t+a_5(x_1)} 
 \qquad \textrm{for each} \quad x\, \epsilon\, (a,b), \\
 \Phi  = C_\Phi \exp{\left[
- \frac{ 4 a_6\int_a^b  u(x)  \drm x +    \int_a^b\int_a^b  
 a_4(x') u(x) u(x')\drm x'   \drm x}
{4 \int_a^b \left( a_4(x)\tau+a_5(x)\right) \drm x} 
\right]}.
\end{eqnarray}
The constants constitute a new set of variables.
We treat $C_\Phi$ as a new dependent variable, hence, from
a relation $C_\Phi=h(\tau,[C(x)])$ we find a formula for
$\Phi$
\begin{eqnarray}
 \Phi  =  h(\tau,[C(x)]) \exp{\left[
- \frac{4 a_6\int_a^b  u(x)  \drm x +   \int_a^b\int_a^b  
 a_4(x') u(x) u(x')\drm x'   \drm x}
{4 \int_a^b \left( a_4(x)\tau+a_5(x)\right) \drm x} 
\right]} \nonumber \\
=  h(\tau,[C(x)]) \exp{\left( H \right) }
\end{eqnarray} 
where for brevity the term in the exponent 
is denoted by $H$.
This result can be further substituted into Eq.~(\ref{h_c})
to find the final form of the invariant solution.
First, we compute the necessary derivatives in terms
of new variables
\begin{eqnarray}\label{tder_ex}
\pd{\Phi}{t} = \pd{h}{\tau}  \exp{\left( H \right) }
+ h \exp{\left( H \right) } H
\frac{\int_a^b  a_4(x) \drm x }
{ \int_a^b \left( a_4(x)\tau+a_5(x)\right) \drm x },
\end{eqnarray}
the expression for the second order functional 
derivative of $\Phi$ is lengthy, 
however, after integration with
respect to $x$ all terms that contain functional
derivatives of $h$ will cancel. Below, we write
the final form, after integration
\begin{eqnarray}\label{funder_ex}
\int_a^b \pdel{^2\Phi}{u(x)^2} \drm x
&=& h \exp{\left( H \right)} \left[
  \frac{2 a_6  +    \int_a^b  
 a_4(x') u(x')\drm x' }
{2 \int_a^b \left( a_4(x)\tau+a_5(x)\right) \drm x}\right]^2 
\nonumber \\
&-& h \exp{\left( H \right)}
  \frac{\int_a^b a_4(x) \drm x  }
{ 2 \int_a^b \left( a_4(x)\tau+a_5(x)\right) \drm x } 
\end{eqnarray}
We can now compare
 (\ref{tder_ex}) with (\ref{funder_ex}) and perform the time
integration to obtain the form of functional $h$.
Finally, the  solution of Eq.\ (\ref{h_c}), invariant
under the considered symmetries writes
\begin{eqnarray}
&\Phi(t,[u(x)]) = \\
&\frac{g([C(x)])}
{\sqrt{t\int_a^b a_4(x)\drm x + \int_a^b a_5(x)  \drm x  }}
\exp{\left[
-\frac{ \left(2 a_6 + \int_a^b a_4(x)u(x)\drm x \right)^2}
{4 \int_a^b a_4(x)\drm x
\left( t\int_a^b a_4(x)\drm x + \int_a^b a_5(x)  \drm x\right)}
  \right]}. \nonumber
\end{eqnarray}

\section{Conclusions and perspectives.}
In the present work the classical, point symmetry group
analysis is extended from partial differential equations to
their counterparts in the continuum limit.
In particular, we introduce the procedure of applying
symmetry analysis to the case when functional derivatives are present in
the equation.
As example the method
is further applied to the continuum limit of a heat equation
and the Lie point symmetries, admitted by this equations 
are derived. 
From the infinitesimal transformations one can also find
the invariant solutions of the considered equations.

The  presented extension of the Lie groups
can be a useful tool for analysing the functional equations.
Though we have only given the heat equation as an
application of the method we  believe that
the new approach is highly relevant to a variety
of important functional differential equations
(FDE) in mathematical physics. Generally
speaking, very little is known on how to
analytically treat and solve FDEs (numerical
treatment is difficult anyway because of the high
dimensionality). Hence, the methods may give a
chance to treat equations which so far have been
put aside because of the missing analytical
methods. In fact, the benefit is twofold since
the symmetries not only allow for analytical
solutions but are also useful in itself since
symmetries illuminate the axiomatic properties of
the physical model equations.

\vskip 1cm
{ \Large Acknowledgements }
\vskip 0.5cm
The authors are thankful to V.\ N.\ Grebenev for his
useful comments and discussions concerning the paper.

\end{document}